%
%
%
%
%
%
%
%
%
%
\hoffset=0.0in
\voffset=0.0in
\hsize=6.5in
\vsize=8.9in
\normalbaselineskip=12pt
\normalbaselines
\topskip=\baselineskip
\parindent=15pt
%
%
%

\let\ra=\rangle

\let\lf=\left
\let\rt=\right

\let\del=\nabla

\let\h=\hbar

\let\rta=\rightarrow

\let\ssy=\scriptscriptstyle
\let\:=\>
\let\\=\cr
\let\emph=\e

\let\m=\hbox

\let\cl=\centerline

\def\e#1{{\it #1\/}}
\def\textbf#1{{\bf #1}}
\def\[{$$}
\def\]{\[}
\def\re#1#2{$$\matrix{#1\cr}\eqno{({\rm #2})}$$}
\def\de#1{$$\matrix{#1\cr}$$}

\def\eqdf{\buildrel{\rm def}\over =}

\def\hfs{{\ssy {1 \over 2}}}

\def\mathrm#1{{\rm #1}}
\def\mr#1{{\rm #1}}
\def\mathcal#1{{\cal #1}}

\def\gip{{\rm GIP}}
\def\Schr{Schr\"{o}\-ding\-er}
\font\frtbf = cmbx12 scaled \magstep1
\font\twlbf = cmbx12
\font\ninbf = cmbx9
\font\svtrm = cmr17
\font\twlrm = cmr12
\font\ninrm = cmr9
\font\ghtrm = cmr8

\def\gr#1{{\ghtrm #1}}

\def\abstract#1{{\ninbf\cl{Abstract}}\medskip
\openup -0.1\baselineskip
{\ninrm\leftskip=2pc\rightskip=2pc\noindent #1\par}
\normalbaselines}
\def\sct#1{\vskip 1.33\baselineskip\noindent{\twlbf #1}\medskip}

\def\so{\raise 0.65ex \m{\sevenrm 1}}
\def\sk{\par\vskip 0.66\baselineskip}
{\svtrm
\cl{The Significance to Quantum Computing of the}
\medskip
\cl{Classical Harmonic Nature of Energy Eigenstates}
}
\bigskip
{\twlrm
\cl{Steven Kenneth Kauffmann\footnote{${}^\ast$}{\gr{Retired, American
Physical Society Senior Life Member, E-mail: SKKauffmann@gmail.com}}}
}
\bigskip\smallskip
\abstract{%
Since a pure quantum system is incapable of faithfully simulating the
solutions of the \Schr\ equation that actually pertains to itself, it
is proposed that quantum computing technology (as opposed to crypto%
graphic technology) not be based on pure quantum systems such as qu%
bits but instead on physical systems which by their nature faithfully
simulate the solutions of \Schr\ equations. Every \Schr\ equation is
within a unitary transformation of being a set of mutually independent
classical simple harmonic oscillator equations.  Thus classical simple
harmonic oscillators, or ``chobits'', are the mathematically fundamen%
tal building blocks for all \Schr\ equations.  In addition, classical
harmonic oscillators are, as a practical matter, far easier to deal
with than any pure quantum system---e.g., their phases and absolute
amplitudes are readily physically accessible, they have little predi%
lection for environmental decoherence, and they abound as cavity elec%
tromagnetic standing-wave modes.  We study in mathematical detail the
use of chobits to compute discrete quantum Fourier transforms, includ%
ing gates, chobit counts, and chobit operation counts.  The results
suggest that thirty chobits and under a thousand chobit phase opera%
tions could generate discrete quantum Fourier transforms of a billion
terms.  Chobits can be technologically realized as semiconductor
dynatron-type electronic oscillator circuits, which ought to be amen%
able to very considerable miniaturization.
}

\sct{Introduction: the \Schr\ equation's classical canonical charac%
ter}
\noindent
The procedures of second quantization~[1] foster awareness that a
\Schr\ equation describes \e{a classical dynamical system in canonical
form}.  This basic but not intuitively expected fact is almost never
pointed out, however, in expositions of quantum mechanics that do not
treat second quantization.  Therefore we now explicitly show that the
complex-valued \Schr\ equation,
\re{
    i\h\dot\psi = H\psi.
}{1a}
for an $M$-state quantum system is equivalent to a purely real-valued
classical canonical equation system.  To this end we define the fol%
lowing two purely real-valued vectors that each have $M$ components,
\re{
    q\eqdf   (\h/2)^\hfs (\psi + \psi^\ast) \quad\mr{and}\quad
    p\eqdf -i(\h/2)^\hfs (\psi - \psi^\ast),
}{1b}
which are such that,
\re{
    \psi = (q + ip)/(2\h)^\hfs.
}{1c}
We also define the following two purely real-valued $M$ by $M$ matri%
ces,
\re{
    H_S\eqdf   (H + H^\ast)/2 =  (H + H^T)/2 \quad\mr{and}\quad
    H_A\eqdf -i(H - H^\ast)/2 = -i(H - H^T)/2,
}{1d}
where the second equality in each of the two parts of Eq.~(1d) follows
from the fact that $H$ is Hermitian, i.e., that $H^\ast = H^T$.
Therefore $H_S$ is a \e{symmetric} matrix as well as being a real-val%
ued one, and $H_A$ is an \e{antisymmetric} matrix as well as being
real-valued.  The definitions of $H_S$ and $H_A$ also imply that,
\re{
    H = H_S + iH_A.
}{1e}
Putting Eqs.~(1c) and (1e) into Eq.~(1a), the \Schr\ equation, produc%
es,
\re{
    -\dot p + i\dot q =  \lf( (H_S/\h)q - (H_A/\h)p \rt) +
                        i\lf( (H_S/\h)p + (H_A/\h)q \rt),
}{1f}
which implies the two purely real-valued equations,
\re{
    \dot q =  (H_S/\h)p + (H_A/\h)q \quad\mr{and}\quad
    \dot p = -(H_S/\h)q + (H_A/\h)p.
}{1g}
It is \e{also} readily verified that Eq.~(1g) together with the defin%
itions given in Eqs.~(1b) and (1d) \e{implies} Eq.~(1a).  Therefore
Eq.~(1g) is \e{equivalent} to Eq.~(1a), the \Schr\ equation.  Now if a
classical Hamiltonian ${\cal H}_{\mr{cl}}^{(H)}(q,p)$ \e{exists} such
that the two equalities given by Eq.~(1g) are the \e{same} as,
\re{
    \dot q =  \del_p {\cal H}_{\mr{cl}}^{(H)}(q, p) \quad\mr{and}\quad
    \dot p = -\del_q {\cal H}_{\mr{cl}}^{(H)}(q, p),
}{1h}
then Eq.~(1g) is a classical dynamical equation system in canonical
form.  It is in fact readily verified, using the facts that $H_S$ is
a real-valued \e{symmetric} matrix and $H_A$ is a real-valued \e{anti%
symmetric} matrix, that the \e{particular} classical Hamiltonian,
\re{
    {\cal H}_{\mr{cl}}^{(H)}(q, p)\eqdf [(q, H_S q) + (p, H_S p) +
                                             2(p, H_A q)]/(2\h),
}{1i}
\e{does indeed} fulfill the condition that Eq.~(1h) is the \e{same} as
Eq.~(1g).  Therefore the generic \Schr\ equation of Eq.~(1a) \e{is in%
deed equivalent to a classical dynamical equation system in canonical
form}.

Finally, if we use the antisymmetric nature of the matrix $H_A$ to
reexpress the classical Hamiltonian of Eq.~(1i) as,
\re{
    {\cal H}_{\mr{cl}}^{(H)}(q, p) = [(q, H_S q) + (p, H_S p) +
                              (p, H_A q) - (q, H_A p)]/(2\h),
}{1j}
and then substitute the definitions given by Eqs.~(1b) and (1d) into
the right-hand side of Eq.~(1j), there results, after a slightly te%
dious gathering and cancellation of terms,
\re{
    {\cal H}_{\mr{cl}}^{(H)}(q, p) = [(\psi, H^\ast\psi^\ast) +
                     (\psi^\ast, H\psi)]/2 = (\psi^\ast, H\psi),
}{1k}
where the last equality follows from the fact that $H^\ast = H^T$.
Therefore \e{the classical Hamiltonian for a \Schr\ equation is equal
to the quantum expectation value of that equation's Hamiltonian ma%
trix}, a result which is very familiar in the context of second
quantization~[1].

While we have so far been dealing with the \e{generic} \Schr\ equa%
tion, there is no real loss of generality from assuming that its Her%
mitian Hamiltonian matrix is \e{diagonal}, because that diagonaliza%
tion can \e{always} in principle be achieved by a \e{unitary transfor%
mation}, which is \e{invertible} and \e{even} of course \e{linear}.
We shall now see that the \Schr\ equation for an $M$-state quantum
system describes, when its Hermitian Hamiltonian matrix is diagonal,
nothing more than $M$ mutually independent classical simple harmonic
oscillators.

\sct{\Schr\ equation simulation by classical simple harmonic oscilla%
tors}
\noindent
When the $M$ by $M$ Hermitian Hamiltonian matrix $H$ of the \Schr\
equation of Eq.~(1a) is \e{diagonal}, then $H = H^T = H^\ast$, and
therefore from Eq.~(1d) we will have that $H_S = H$ and $H_A = 0$,
which simplifies the classical dynamical equations of Eq.~(1g) to,
\re{
    \dot q =  (H/\h)p \quad\mr{and}\quad \dot p = -(H/\h)q.
}{2a}
Furthermore, the \e{diagonal} $M$ by $M$ Hermitian Hamiltonian matrix
$H$ will satisfy,
\re{
    H|m\ra = E_m|m\ra, \quad m = 0, 1, \ldots, M - 1,
}{2b}
where $|0\ra, |1\ra, \ldots, |M - 1\ra$ are the natural orthonormal
basis vectors for the $M$-dimensional vector space which have a single
sequentially selected component set equal to unity with the rest set
equal to zero, and $E_0, E_1, \ldots, E_{M-1}$ are \e{the correspond%
ing diagonal elements (i.e., energy eigenvalues) of} $H$.  We can use
this natural complete orthonormal basis set to decompose the vectors
$q$ and $p$ into their components,
\re{
    q = \sum_{m = 0}^{M - 1} q_m |m\ra \quad\mr{and}\quad
    p = \sum_{m = 0}^{M - 1} p_m |m\ra.
}{2c}
Eq.~(2c) in conjunction with both Eqs.~(2a) and (2b) implies that,
\re{
    \dot q_m =  (E_m/\h)p_m \enskip\mr{and}\enskip
    \dot p_m = -(E_m/\h)q_m, \enskip m = 0, 1, \ldots, M - 1,
}{2d}
which we recognize as the classical dynamical equations of $M$ \e{mu%
tually independent classical simple harmonic oscillators}.  We can
likewise decompose the complex-valued vector $\psi$ into its complex-%
valued components,
\re{
    \psi = \sum_{m = 0}^{M - 1} \psi_m |m\ra,
}{2e}
which in conjunction with Eq.~(2b) and the \Schr\ equation given by
Eq.~(1a) implies that,
\re{
    \dot \psi_m = -i(E_m/\h)\psi_m, \enskip m = 0, 1, \ldots, M - 1.
}{2f}
These $M$ independent differential equations have the general solu%
tions,
\re{
    \psi_m(t) = \psi_m(t_0)\exp(-iE_m(t - t_0)/\h),
    \enskip m = 0, 1, \ldots, M - 1.
}{2g}
Eqs.~(2g) and (2e) yield the general solution to the \Schr\ equation
given by Eq.~(1a), namely,
\re{
    \psi(t) = \sum_{m = 0}^{M - 1} \psi_m(t_0)
              \exp(-iE_m(t - t_0)/\h)|m\ra.
}{2h}
Eq.~(2h) is, in conjunction with Eq.~(2b), readily shown to indeed sa%
tisfy the \Schr\ equation given by Eq.~(1a).  Furthermore, Eq.~(2h) in
conjunction with Eq.~(1b) implies that,
\re{
    q(t) = (2\h)^\hfs\sum_{m = 0}^{M - 1}|\psi_m(t_0)|
           \cos\bigl(\arg(\psi_m(t_0)) - E_m(t - t_0)/\h\bigr)|m\ra,
    \cr
    p(t) = (2\h)^\hfs\sum_{m = 0}^{M - 1}|\psi_m(t_0)|
           \sin\bigl(\arg(\psi_m(t_0)) - E_m(t - t_0)/\h\bigr)|m\ra,
}{2i}
which, in conjunction with Eq.~(2b), is readily shown to indeed satis%
fy the classical dynamical equations given by Eq.~(2a).  From Eq.%
~(2i), which precisely corresponds to Eq.~(2c), we readily isolate the
fully solved dynamics of the $M$ mutually independent classical simple
harmonic oscillators whose classical dynamical equations are given by 
Eq.~(2d),
\re{
    q_m(t) = (2\h)^\hfs|\psi_m(t_0)|
           \cos\bigl(\arg(\psi_m(t_0)) - E_m(t - t_0)/\h\bigr),
    \cr
    p_m(t) = (2\h)^\hfs|\psi_m(t_0)|
           \sin\bigl(\arg(\psi_m(t_0)) - E_m(t - t_0)/\h\bigr),
}{2j}
where $m = 0, 1, \ldots, M - 1$.  The $q_m(t)$ and $p_m(t)$ of Eq.~%
(2j) are readily shown to indeed satisfy the classical dynamical equa%
tions that are given by Eq.~(2d).  These $M$ mutually independent
classical simple harmonic oscillator solutions furthermore have \e{ex%
actly} the required absolute amplitudes, namely $(2\h)^\hfs|\psi_m(t_0
)|$, \e{and} the required phases, namely $\arg(\psi_m(t_0))$, \e{to
precisely simulate the general solution} $\psi(t)$ given by Eq.~(2h)
to the \Schr\ equation for an $M$-state quantum system that is given
by Eq.~(1a).  Indeed, combining Eq.~(2j) with Eqs.~(2c) and (1c)
yields precisely Eq.~(2h).

Thus we see that the solutions of \Schr\ equations for $M$-state quan%
tum systems \e{can always be faithfully simulated by} $M$ \e{mutually
independent classical simple harmonic oscillators}.  We therefore now
dub the \e{most basic} classical dynamical entity for the faithful
simulation of \Schr\ equations the ``chobit'', and note that the
``chobit'' is, of course, \e{a single classical simple harmonic os%
cillator}---the prefix ``cho'' abbreviates ``classical harmonic oscil%
lator''.  The chobit, given its amplitude and phase, represents \e{a
single complex number}.  A complex number of course comprises far more
data than the integer modulo two represented by an ordinary bit.

It is very worthwhile to take notice at this point of the fact that
the \e{physical quantum system} to which a given \Schr\ equation
\e{pertains}, in principle \e{cannot faithfully simulate the solutions
of that \Schr\ equation} because the \e{quantum mechanics} of a physi%
cal system \e{is not in one-to-one correspondence with the solutions
of the \Schr\ equation that pertains to it}.  For example, some of the
information that is \e{an inherent part} of \Schr-equation solutions
is systematically \e{lost} because of the probabilistic requirement
that physical meaning be attached to \e{only the absolute squares} of
inner products of state vectors---this makes the \e{inherent overall
phase} of any state vector \e{physically meaningless}.  \e{Further}
information that is \e{an inherent part} of \Schr-equation solutions
is lost because of the probabilistic requirement that state vector
\e{norms} be \e{devoid of physical meaning}.  In fact, \e{single}-%
state quantum systems, which are described by the simplest \Schr\
equations that are possible (and are simulated by a single chobit),
have \e{all} of their \Schr-equation solution information, namely
\e{both} their \e{single} phase \e{and} their \e{single} absolute
amplitude, \e{made physically meaningless} by these two probabilistic
requirements.  Consequently, \e{single}-state quantum systems \e{do
not physically exist at all}.  Physical \e{two}-state quantum systems,
the celebrated qubits, whose \Schr\ equations require \e{two} chobits
to simulate, i.e., \e{two complex numbers}, have \e{a very significant
fraction of that information made physically meaningless} by these
probabilistic requirements---this has the consequence that physical
qubits can bear information equivalent to \e{only two angles}, one
azimuthal and one polar.  Thus a physical qubit \e{falls drastically
short} of being able to faithfully simulate the solutions of the
two-chobit \Schr\ equation which pertains to it.

Furthermore, even information that \e{is} borne by a quantum system
can sometimes be so awkward to recover as to be poorly suited to com%
putational applications.  For example, the superposition of orthonor%
mal basis states with equal absolute amplitudes but differing phases
is annoyingly recalcitrant with regard to recovery of the relative
phase information: the simple inner product of that superposition
state with any one of the basis states results in the \e{complete
obliteration of the phase information} by the probabilistically manda%
tory subsequent taking of the \e{absolute square} of such an inner
product.  For computational applications chobits thus absolutely shine
by comparison with actual quantum systems such as physical qubits be%
cause the \e{classical nature of chobits} in principle makes \e{all
the information they bear readily physically accessible}.  (But by the
very same token \e{actual} physical \e{quantum} systems offer superior
potential for ingenious \e{cryptography}.)

The fact that chobit \e{phases} are in principle readily physically
accessible is of \e{particular relevance} to carrying out the uni%
tary \e{discrete quantum Fourier transform}~[2] whose  definition is,
\re{
    U^{(M)}_F|m\ra\eqdf M^{-\hfs}\sum_{m' = 0}^{M - 1}
                        e^{i2\pi m'm/M}|m'\ra \enskip\mr{where}
                        \enskip m\in\{0, 1, \ldots, M - 1\}.
}{3a}
The right-hand side of Eq.~(3a) is \e{precisely} the sort of superpo%
sition of orthonormal basis states with equal absolute amplitudes but
differing phases mentioned above that, for \e{actual} physical
\e{quantum} systems, makes recovery of the relative phase information
so awkward.  In the next section, where we discuss Eq.~(3a) in detail,
it is therefore \e{to be implicitly understood that the phases are
represented by chobits}, and this \e{still applies} after Eq.~(3a) is
recast into computationally advantageous direct-product forms~[2].

\sct{The chobit discrete quantum Fourier transform}
\noindent
The discussion below of the discrete quantum Fourier transform that is
defined by Eq.~(3a) \e{closely parallels} that given in Ref.~[2], but
the adoption of the chobit in conjunction with the jettisoning of any
thought of using actual physical qubits invites the exploration of the
obvious $p$-nary \e{generalizations} of the \e{binary} direct-product
representation of Eq.~(3a) that is presented in Ref.~[2].

For some arbitrary integer $p\geq 2$ we take the $M$ of Eq.~(3a) to be
\e{equal} to $p^n$, where one normally expects the positive integer
$n$ to be much greater than unity.  Since the integer $m'$ in Eq.~(3a)
satisfies $0\leq m'\leq p^n - 1$, we can write the $p$-nary expansion
of $m'$ as,
\re{
    m' = \sum_{l = 1}^n m'_l p^{n - l},
}{3b}
where the integers $m'_l$ satisfy $m'_l\in\{0, 1, \ldots, p - 1\}$ for
$l = 1, 2, \ldots, n$.  This expansion permits us to write the integer
$m'$ as its $n$-digit $p$-nary representation $m' = m'_1m'_2\ldots
m'_n$.  This $n$-digit $p$-nary representation of $m'$ in turn permits
us to express the basis state $|m'\ra$ of the $p^n$-state system as an
$n$-fold \e{direct product} of basis states of $p$-state systems,
\re{
    |m'\ra = |m'_1\ra\otimes|m'_2\ra\otimes\ldots\otimes|m'_n\ra =
             \otimes_{l = 1}^n|m'_l\ra.
}{3c}
In analogous fashion, the integer $m$ in Eq.~(3a), which satisfies $0
\leq m\leq p^n - 1$, is reexpressed by its $p$-nary expansion as,
\re{
    m = \sum_{l = 1}^n m_l p^{n - l} = \sum_{r = 1}^n m_{n + 1 -r} p^{r - 1},
}{3d}
where the integers $m_l$ satisfy $m_l\in\{0, 1, \ldots, p - 1\}$ for
$l = 1, 2, \ldots, n$.  Thus, exactly as for the integer $m'$ above,
we can write the integer $m$ as its $n$-digit $p$-nary representation
$m = m_1m_2\ldots m_n$, which in turn permits us to express the basis
state $|m\ra$ of the $p^n$-state system as an $n$-fold \e{direct pro%
duct} of basis states of $p$-state systems,
\re{
    |m\ra = |m_1\ra\otimes|m_2\ra\otimes\ldots\otimes|m_n\ra =
             \otimes_{l = 1}^n|m_l\ra.
}{3e}
We shall be using Eq.~(3e) for the basis state $|m\ra$ which appears
on the left-hand side of Eq.~(3a), but for the $m$ which appears in
the exponent on the right-hand side of Eq.~(3a) we shall in due course
be using the rightmost summation representation given in Eq.~(3d).

In light of Eq.~(3b), we now replace the factors $m'/M = m'/p^n$ in
the exponent on the right-hand side of Eq.~(3a) by $\sum_{l = 1}^n
m'_l p^{-l}$.  In light of Eq.~(3c) we also replace $|m'\ra$ on the
right-hand side of Eq.~(3a) by $\otimes_{l = 1}^n|m'_l\ra$.  To pro%
perly \e{match} these decompositions involving $m'$, we note that the
following \e{notational changes} must \e{also} be carried out on the
right-hand side of Eq.~(3a),
\de{
    M^{-\hfs}\sum_{m' = 0}^{M - 1}\rta
    (p^n)^{-\hfs}\sum_{m' = 0}^{p^n - 1}\rta
    p^{-\hfs}\sum_{m'_1 = 0}^{p - 1}\cdots 
    p^{-\hfs}\sum_{m'_n = 0}^{p - 1}.
}
With these changes and a subsequent convenient rearrangement of fac%
tors, Eq.~(3a) becomes,
\re{
    U^{(p^n)}_F|m\ra = \otimes_{l = 1}^n\lf[
                     p^{-\hfs}\sum_{m'_l = 0}^{p - 1}
                     e^{i2\pi m'_l p^{-l}m}|m'_l\ra\rt]
                     \enskip\mr{where}\enskip
                     m\in\{0, 1, \ldots, p^n - 1\}.
}{3f}
We now note from Eq.~(3e) that $|m\ra = \otimes_{l = 1}^n|m_l\ra$,
where each $m_l\in\{0, 1, \ldots, p - 1\}$ for $l = 1, 2, \ldots, n$.
We also note from the rightmost summation representation given in
Eq.~(3d) that $p^{-l}m = \sum_{r = 1}^n m_{n + 1 -r}p^{r - l - 1}$,
and that all the \e{terms} of this sum for which $r\geq l + 1$ have
\e{integer values}.  Therefore Eq.~(3f) becomes,
\re{
    U^{(p^n)}_F\bigl(\otimes_{l = 1}^n|m_l\ra\bigr) =
                     \otimes_{l = 1}^n\lf[
                     p^{-\hfs}\sum_{m'_l = 0}^{p - 1}
                     e^{i2\pi m'_l\sum_{r = 1}^l
                     m_{n + 1 - r}p^{r - l - 1}}
                     |m'_l\ra\rt],
}{3g}
which is more transparent when expressed in terms of ascending powers
of $p^{-1}$,
\re{
    U^{(p^n)}_F\bigl(\otimes_{l = 1}^n|m_l\ra\bigr) =
                     \otimes_{l = 1}^n\lf[
                     p^{-\hfs}\sum_{m'_l = 0}^{p - 1}
                     e^{i2\pi m'_l\sum_{s = 1}^l
                     m_{n - l + s}p^{-s}}
                     |m'_l\ra\rt].
}{3h}
By using ``$p$-nary point'' notation, Eq.~(3h) can alternatively be
written,
\re{
    U^{(p^n)}_F\bigl(\otimes_{l = 1}^n|m_l\ra\bigr) =
                     \otimes_{l = 1}^n\lf[
                     p^{-\hfs}\sum_{m'_l = 0}^{p - 1}
                     e^{i2\pi m'_l(0.m_{n - l + 1}
                     m_{n - l + 2}\ldots m_n)}
                     |m'_l\ra\rt],
}{3i}
Note that when $n = 1$ one recovers from Eq.~(3h), (3i) or (3g) the
original Eq.~(3a) with $M = p$, $m = m_1$, and $m' = m'_1$.

The salient point of the discrete quantum Fourier transform $p$\e{-%
nary direct-product representations} of Eqs.~(3h),(3i) or (3g) is that
they feature only $n$ times $p-1$ phases that require chobits, whereas
the \e{original} discrete quantum Fourier transform representation of
Eq.~(3a) features $M-1 = p^n-1$ phases that require chobits.  An in%
teresting small exercise is to hold $p^n$, the total number of terms
of the discrete quantum Fourier transform  (which is related to the
resolution achieved by that discrete Fourier transform), \e{fixed}
while simultaneously attempting to \e{minimize} $n(p-1)$, the number
of chobits needed to accommodate the number of phases in the $p$-nary
direct product representation of that discrete quantum Fourier trans%
form.  Since $n(p-1) = ln(p^n)[(p-1)/ln(p)]$, we need to \e{minimize}
$(p-1)/ln(p)$, a function of $p$ that \e{increases monotonically when}
$p > 0$.  Therefore the $p=2$ binary base version of the direct-pro%
duct representation of the discrete quantum Fourier transform that is
presented in Ref.~[2] \e{minimizes} the number of chobits needed for a
given number of terms of that transform.  With $p=2$, $n$ chobits suf%
fice for $2^n$ terms.

Ref.~[2] shows how the particular $p=2$ \e{binary} version of the di%
rect-product representation of the discrete quantum Fourier transform
given by Eq.~(3i) or (3h) can be built up from interwoven repetitions
of a few elementary unitary \e{gates}, namely operations that are each
based on data stored in a \e{single} (binary) digit.  The small number
of unitary \e{single binary-digit} gates which are so utilized in
Ref.~[2] can be all be \e{straightforwardly generalized} into unitary
\e{single} $p$\e{-nary-digit} gates.  The $p$-nary generalization of
the unitary ``Hadamard gate'' of Ref.~[2] is an especially interesting
example of such a gate generalization.  Given a $p$-nary \e{single-%
digit} state $|m_l\ra$, where $m_l\in\{0, 1, \ldots, p - 1\}$ and $l
\in\{1, 2, \ldots, n\}$, the application to it of the unitary ``gener%
alized Hadamard gate'' \e{changes this one-digit state} $|m_l\ra$ into
the particular \e{one-digit superposition state},

\re{
    p^{-\hfs}\sum_{m'_{n + 1 - l} = 0}^{p - 1}
    e^{i2\pi m'_{n + 1 - l}m_l/p}|m'_{n + 1 - l}\ra.
}{4a}
Comparing $|m_l\ra$ and the one-digit superposition state given by
Eq.~(4a) with the two sides of Eq.~(3a), we see that the unitary
``generalized Hadamard gate'' is a microcosmic \e{one-digit analog}
of the unitary discrete quantum Fourier transform \e{itself}.

We proceed now to build up Eq.~(3h) in very close analogy with the de%
tailed interwoven repetitions of \e{gate applications} presented in
Ref.~[2].  Starting with the $n$\e{-digit direct-product state} $|m\ra
= |m_1\ra\otimes\ldots\otimes|m_n\ra = \otimes_{l=1}^n|m_l\ra$ that
appears on the left-hand side of Eq.~(3h), we commence by applying the
``generalized Hadamard gate'' of Eq.~(4a) to its \e{leftmost digit}
$|m_1\ra$, which yields,
\re{
     \lf[p^{-\hfs}\sum_{m'_n = 0}^{p - 1}
     e^{i2\pi m'_nm_1/p}|m'_n\ra\rt]
     \otimes\bigl(\otimes_{l=2}^n|m_l\ra\bigr).
}{4b}
Note from Eq.~(4b) that the effect of the generalized Hadamard gate
on the leftmost digit $|m_1\ra$ has been to \e{change} it into a one-%
digit superposition state which is a one-digit analog of the discrete
quantum Fourier transform itself, with the digit $|m_2\ra$ \e{remain%
ing in place} to become the \e{successor} leftmost digit, i.e., the
generalized Hadamard gate has, inter alia, effectively \e{consumed}
the leftmost digit $|m_1\ra$, leaving $|m_2\ra$ behind as the leftmost
\e{available} digit.

We next need to \e{modify} the one-digit superposition state that the
generalized Hadamard gate has just created with \e{the information
that is stored in the remaining available digits} $|m_2\ra, |m_3\ra,
\ldots, |m_n\ra$, and we achieve that end with a \e{cascading
sequence} of gates that are closely similar to each other.  Those
remaining available digits \e{themselves}, however, \e{are in no way
modified} and therefore \e{are not consumed} by this gate cascade.

We begin the gate cascade by applying the generalized $R_2$ gate~[2]
to the leftmost available digit $|m_2\ra$ in Eq.~(4b) in order to in%
sert that digit's information into the one-digit superposition state
that was created by the preceding generalized Hadamard gate.  This
changes Eq.~(4b) to,
\re{
     \lf[p^{-\hfs}\sum_{m'_n = 0}^{p - 1}
     e^{i2\pi m'_n[m_1/p + m_2/p^2]}|m'_n\ra\rt]
     \otimes\bigl(\otimes_{l=2}^n|m_l\ra\bigr).
}{4c}
We now continue with the rest of the generalized $R_k$ gate~[2] cas%
cade by applying the similar generalized $R_3$ gate~[2] to the $|m_3
\ra$ digit in order to insert that digit's information into the one-%
digit superposition state that was created by the preceding general%
ized Hadamard gate, then applying the generalized $R_4$ gate~[2] to
the $|m_4\ra$ digit in order to do the same with that digit's informa%
tion, and so forth, finally applying the generalized $R_n$ gate~[2] to
the $|m_n\ra$ digit in order to do so with that digit's information.
The result of this particular \e{entire} generalized $R_k$ gate~[2]
\e{cascade} applied to Eq.~(4b) is,
\re{
     \lf[p^{-\hfs}\sum_{m'_n = 0}^{p - 1}
     e^{i2\pi m'_n\sum_{s = 1}^n m_s p^{-s}}|m'_n\ra\rt]
     \otimes\bigl(\otimes_{l=2}^n|m_l\ra\bigr).
}{4d}
Having \e{finished} this particular generalized $R_k$ gate~[2] \e{cas%
cade}, which has \e{not} consumed \e{any} of the digits that were
available in Eq.~(4b), we now apply the generalized Hadamard gate to
the leftmost available digit in Eq.~(4d), namely $|m_2\ra$, which
\e{consumes that digit} and produces a \e{second} one-digit superposi%
tion state, so that Eq.~(4d) becomes,
\re{
     \lf[p^{-\hfs}\sum_{m'_n = 0}^{p - 1}
     e^{i2\pi m'_n\sum_{s = 1}^n m_s p^{-s}}|m'_n\ra\rt]\otimes
     \lf[p^{-\hfs}\sum_{m'_{n - 1} = 0}^{p - 1}
     e^{i2\pi m'_{n - 1}m_2/p}|m'_{n - 1}\ra\rt]
     \otimes\bigl(\otimes_{l=3}^n|m_l\ra\bigr).
}{4e}
Now we \e{modify} this \e{second} one-digit superposition state which
has been created by the \e{latest} generalized Hadamard gate with
\e{another} generalized $R_k$ gate~[2] \e{cascade}.  These generalized
$R_k$ gate~[2] cascades \e{always begin} with the application of the
generalized $R_2$ gate~[2] to the \e{leftmost available digit} which
here is, from Eq.~(4e), $|m_3\ra$.  This generalized $R_k$ gate cas%
cade continues from there, with the generalized $R_3$ gate applied to
$|m_4\ra$, the generalized $R_4$ gate applied to $|m_5\ra$ and so
forth, finally ending with the generalized $R_{n - 1}$ gate applied to
$|m_n\ra$.  The upshot of this generalized $R_k$ gate \e{cascade},
which \e{modifies} the one-digit superposition state that was created
by the \e{most recent} generalized Hadamard gate, is to change Eq.~%
(4e) to,
\re{
     \lf[p^{-\hfs}\sum_{m'_n = 0}^{p - 1}
     e^{i2\pi m'_n\sum_{s = 1}^n m_s p^{-s}}|m'_n\ra\rt]\otimes
     \lf[p^{-\hfs}\sum_{m'_{n - 1} = 0}^{p - 1}
     e^{i2\pi m'_{n - 1}\sum_{s = 1}^{n - 1}
     m_{s + 1} p^{-s}}|m'_{n - 1}\ra\rt]
     \otimes\bigl(\otimes_{l=3}^n|m_l\ra\bigr),
}{4f}
which, of course, has \e{not} consumed \e{any} of the digits that were
available in Eq.~(4e).

We continue in this way with the application of the generalized Hada%
mard gate to the leftmost available digit (in Eq.~(4f) that would be
the digit $|m_3\ra$), which consumes that digit, followed by a gener%
alized $R_k$ gate \e{cascade} that begins with application of the gen%
eralized $R_2$ gate to the leftmost available digit and ends after ap%
plication of a generalized $R_k$ gate to the rightmost available dig%
it, which is $|m_n\ra$, followed by the application of yet another
generalized Hadamard gate to the leftmost available digit, which con%
sumes that digit, etc.  This procedure finally produces, after the
very last available digit $|m_n\ra$ is consumed by a generalized Hada%
mard gate,
\re{
                     \otimes_{l = n}^1\lf[
                     p^{-\hfs}\sum_{m'_l = 0}^{p - 1}
                     e^{i2\pi m'_l\sum_{s = 1}^l
                     m_{n - l + s}p^{-s}}
                     |m'_l\ra\rt],
}{4g}
which has its direct product factors in the \e{reverse order} to those
of Eq.~(3h).  As is explained in Ref.~[2], \e{swap operations}, which
involve additional gates, are then applied to turn the order of these
reversed factors around, which produces Eq.~(3h).

Exactly as in the straightforward \e{gate census} described in Ref.~%
[2], we readily see that there are altogether $n(n + 1)/2$ generalized
$R_k$ and Hadamard gates, plus $n/2$ swap operations.  Each general%
ized $R_k$ or generalized Hadamard gate requires $p - 1$ \e{changes in
phases} or \e{initializations of phases}.  Therefore the \e{total num%
ber of chobit operations required} is of order $(n(p-1))^2$, i.e., it
is of the  order of the \e{square} of the total number $n(p-1)$ of
chobits required.  Again holding the total number of terms $p^n$ of
the discrete  quantum Fourier transform \e{fixed} while simultaneously
attempting to \e{minimize} the total required number of chobit opera%
tions $(n(p-1))^2$, we note that $(n(p-1))^2 = (ln(p^n))^2[(p-1)/ln(p)
]^2$, and $[(p-1)/ln(p)]^2$ \e{also increases monotonically for} $p >
0$.  Therefore the $p=2$ binary base version of the direct-product re%
presentation of the discrete quantum Fourier transform that is pre%
sented in Ref.~[2] \e{minimizes} the number of chobit operations need%
ed for a given number of terms of that transform.  With $p = 2$, $n$
chobits and of order $n^2$ chobit operations are sufficient for $2^n$
terms.

It is interesting to note at this point that the direct-product repre%
sentation of the discrete quantum Fourier transform that is given by
Eq.~(3h) or (3i) has, from Eq.~(3f), the \e{equivalent} but more sug%
gestive form,
\re{
    U^{(p^n)}_F|m\ra = \otimes_{l = 1}^n\lf[
                     p^{-\hfs}\sum_{m'_l = 0}^{p - 1}
                     e^{i2\pi m'_l(m/p^l -[m/p^l]_{\mr{GIP}})}
                     |m'_l\ra\rt]\enskip\mr{where}\enskip
                     m\in\{0, 1, \ldots, p^n - 1\},
}{5}
and where $[\quad]_\gip$ denotes the ``greatest integer part'' of the
argument enclosed by its square brackets.  Now \e{instead of using the
very complicated interwoven repetitions of gates} set out in Eqs.~(4)
to build up the phases of the direct-product representation, which is
the approach taken in Ref.~[2], we can \e{very simply} develop, in
$p$-nary multiple-precision \e{floating point representation}, the $n$
successive numbers $m/p, m/p^2, \ldots, m/p^n$ that Eq.~(5) needs
through $n$ \e{successive divisions by} $p$ which can be achieved by
\e{mere subtractions of unity from the} $p$\e{-nary floating-point ex%
ponent}, and we can \e{simultaneously} very simply develop the \e{ac%
companying} $n$ successive $p$-nary multiple-precision \e{fixed-point}
integers $[m/p]_\gip, [m/p^2]_\gip, \ldots, [m/p^n]_\gip$ that Eq.~(5)
needs through calculations of $[m/p^l]_\gip$ from $[m/p^{l-1}]_\gip$
that eliminate the latter's \e{least significant} $p$-nary digit via
a \e{mere single-digit rightward shift}.  Since, as has been noted
above, the \e{optimum value} of $p$ is $2$, we are \e{in fact} here
talking about merely the familiar \e{binary} multiple-precision float%
ing-point and fixed-point representations.  Moreover, since $2^{30}
\approx 10^9$, the values of $n$ that are required in practice
\e{likely permit mere double-precision or even single-precision binary
representation} of the $n$ successive number pairs which are based on
the integer $m$ that Eq.~(5) needs.

The straightforward simplicity of the procedure just outlined con%
trasts sharply with the convoluted complexity of the standard applica%
tion of quantum gates set out in Eqs.~(4) and in Refs.~[2] and [3].
Such simplicity has an interesting echo in earlier work on the dis%
crete quantum Fourier transform by R. B. Griffiths and C. S. Niu~[4],
who pointed out that the conversion of intermediate results of quantum
gate operations into \e{classical signals} that control subsequent
quantum gate operations permits considerably fewer and simpler such
gates.  That moving \e{away} from the use of quantum hardware can in
fact \e{increase} computational effectiveness is, of course, \e{not
fortuitous happenstance}---an $M$-state physical quantum system is
sufficiently ill-suited to computation that it cannot simulate the
full solution space of the \Schr\ equation which actually pertains to
\e{itself}.  The inherently awkward, complex and convoluted character%
istics of quantum hardware that ill suit it to straightforward compu%
tation obviously \e{do} provide, however, marvelously fertile ground
for the development of \e{cryptography}.

\sct{Conclusion}
\noindent
Chobit-based computing automatically and naturally removes the awkward
bottlenecks which are \e{inherent} to attempts to base computing tech%
nology on \e{actual} quantum systems such as \e{physical} qubits.
Whereas the latter are by their very nature \e{incapable} of faithful%
ly simulating the full solution space of the very \Schr\ equations
which actually pertain to them, chobits, which \e{themselves} faith%
fully simulate the full set of one-state-system \Schr-equation solu%
tions, are \e{the fundamental building blocks} for simulating \e{all}
\Schr-equation solutions: $M$ chobits suffice to faithfully simulate
the full solution sets of \Schr\ equations which pertain to $M$-state
quantum systems.

The \e{inherent information content} of the chobit is the \e{complex
number}, which could hardly be more convenient for an immense variety
of computing applications.  In the starkest possible contrast, the
probabilistic nature of quantum mechanics deprives the \e{actual quan%
tum counterpart of the chobit} of even an iota of information content,
so that \e{such a one-state quantum system cannot physically exist}.
Even qubits, which when conceptualized \e{within} the \Schr-equation
domain are simply two-chobit systems, are in quantum physical reality
bereft of a considerable fraction of that information content: instead
of two complex numbers the qubit's information content is two angles,
one azimuthal and the other polar, and as is well known, the probabil%
istic nature of quantum mechanics can make even that diminished infor%
mation capacity of the qubit extremely awkward to access in practice-%
--it is indeed the Byzantine nature of quantum information which gives
it such great promise for cryptographic applications.

The chobit, which is a classical simple harmonic oscillator, in con%
trast should in principle present no issues at all with regard to ac%
cess to the phase and absolute amplitude information it carries.  Cho%
bits, being classical systems, in addition do not suffer from the en%
vironmental decoherence issues that are a natural aspect of many ul%
tra-microscopic pure quantum systems.

One possible technological realization of the chobit is a ``dynatron-%
type'' electronic oscillator circuit~[5], in which a powered ``nega%
tive-resistance'' element is placed in parallel with a basic electron%
ic oscillator circuit in order to cancel out that circuit's innate
electrical resistance, thus enabling it to indefinitely sustain simple
harmonic current oscillation at its natural frequency.  If this active
``negative-resistance'' element is an appropriate \e{semiconductor de%
vice} such as a tunnel diode~[6] or Gunn diode~[7], such a ``dynatron%
-type'' electronic oscillator circuit ought to be amenable to very
considerable \e{miniaturization}.  One can therefore envision such
``chobit circuits'' eventually being routinely incorporated into the
designs of large-scale integrated circuits that are intended for ap%
plications in which chobits can be useful.

Gunn diodes, which can operate at higher power levels than tunnel di%
odes, are frequently used to provide active negative resistance for
\e{electromagnetic cavity} oscillators~[7].  The multitudinous elec%
tromagnetic \e{standing-wave modes} of such a cavity could in princi%
ple comprise \e{a very large number of chobits}, but it would be chal%
lenging to accurately detect and manipulate the phases and absolute
amplitudes of such a large set of standing-wave modes.  Nonlinear de%
vices which mix external wave signals with internally generated refer%
ence frequencies to produce ``beat'' frequencies, a technique known in
radio engineering parlance as ``heterodyning'', could at least in
principle be the key to carrying out such a task~[8].

The above considerations concerning conceivably multitudinous chobits
notwithstanding, is is worth noting that the beauty of the direct-%
product representation of the discrete quantum Fourier transform is
that a modest number of chobits suffices to carry that transform out.
For example, thirty chobits suffice for a discrete quantum Fourier
transform of a billion terms, and forty chobits will do for a trillion
terms.

\vfil
\break
\noindent{\frtbf References}
%
\vskip 0.25\baselineskip

{\parindent = 15pt
\sk\item{[1]}
S. S. Schweber,
\e{An Introduction to Relativistic Quantum Field Theory}
(Harper \& Row, New York, 1961).
\sk\item{[2]}
M. A. Nielsen and I. L. Chuang,
\e{Quantum Computation and Quantum Information}
(Cambridge University Press, Cambridge, 2000).
\sk\item{[3]}
P. W. Shor,
SIAM J. Sci.\ Statist.\ Comput.\ {\bf 26},
1484 (1997).
\sk\item{[4]}
R. B. Griffiths and C. S. Niu,
Phys.\ Rev.\ Lett.\ {\bf 76},
3228 (1996).
\sk\item{[5]}
Wikipedia,
``Dynatron oscillator'',
http://en.wikipedia.org/wiki/Dynatron\_oscillator.
\sk\item{[6]}
Wikipedia,
``Tunnel diode'',
http://en.wikipedia.org/wiki/Tunnel\_diode.
\sk\item{[7]}
Wikipedia,
``Gunn diode'',
http://en.wikipedia.org/wiki/Gunn\_diode.
\sk\item{[8]}
Wikipedia,
``Heterodyne'',
http://en.wikipedia.org/wiki/Heterodyne.
}
\bye